\documentclass{ws-procs9x6}

\begin{document}

\title{Spin effects in  large rapidity neutral pion production at STAR}

\author{A. Ogawa for the STAR collaboration}

\address{Brookhaven National Laboratory, \\
Upton, NY 11973-5000, USA\\ 
E-mail: akio@bnl.gov}

\maketitle

\abstracts{
First measurements of large Feynman $x$ ($x_F$) neutral pion production in
polarized proton collisions at $\sqrt{s}=200$ GeV were reported previously.
Cross section measurements at $<\eta>=3.3$ and $3.8$ and analyzing power ($A_N$)
measurements at $<\eta>=3.8$ were completed for $0.3 < x_F < 0.6$.
The cross section was found to be consistent with next-to-leading order
perturbative QCD calculations.  In that framework, the dominant subprocess
is partonic collisions of valence quarks on low Bjorken-$x$ gluons.
The $A_N$ was found to be large and positive and increasing with $x_F$, 
and consistent with phenomonological calculations based on the Collins effect, 
the Sivers effect and initial-state and final-state higher twist contributions.
Subsequent data have been acquired with transverse polarized proton collisions 
at RHIC at $\sqrt{s}=200$ GeV. The status of the analysis and preliminary
results will be presented.
}


In the perturbative QCD picture, high $x_F$ hadron production in 
hadron-hadron collisions probes asymmetric partonic collisions. It is
dominated by the collisions of a large-$x$ quark on a low-$x$ gluon. 
Since the longitudinal polarization of large-$x$ quark is known to be 
large from polarized DIS experiments, this is an interesting place to 
study spin effects in the nucleon. Also it is an ideal probe for
low-$x$ gluons because quarks directly couple with gluons.

An important question to address is whether fixed-order pQCD is
appropriate to describe forward particle production in p+p collisions
at $\sqrt{s}$=200 GeV.  For $\sqrt{s}\leq 62$ GeV, next-to-leading
order pQCD severely underpredicts measured $\pi^0$ cross sections
\cite{soffer}.  At $\sqrt{s}$=200 GeV and larger collision energies,
there is quantitative agreement between NLO pQCD calculations and
measured cross sections at mid-rapidity\cite{phenix}. This agreement
has been found to extend to $\pi^0$ production at $\langle \eta
\rangle=3.8$ and $3.3$\cite{starfpd,eta3.3}.  

Further tests of the underlying dynamics responsible for forward particle
production can be obtained from the study of particle correlations.
In particular, strong azimuthal correlations of hadron pairs are expected 
when particle production arises from $2 \rightarrow 2$ parton scattering.
The azimuthal correlations of hadron pairs separated by large $\Delta\eta$
were reported\cite{dis04}, and found to be consistent with PYTHIA predictions
\cite{pythia}.


In polarized proton-proton collisions, the transverse single spin asymmetry 
($A_N$) for the $p_\uparrow+p\rightarrow\pi^0$+{\it X} reaction has been 
measured\cite{e704,starfpd} at $\sqrt{s}=20$ and $200$ GeV and found to 
be large. Semi-inclusive deep-inelastic lepton scattering experiments 
\cite{hermes,SMCconference} have reported measurements of transverse 
single-spin asymmetries which are significantly different from zero. 

In perturbative QCD at leading twist with collinear factorization, 
$A_N$ has to be zero. If one takes a step beyond this simple scheme and allows
partons to have transverse momentum ($k_T$), there are 3 terms which can contribute 
to the asymmetry.   The first is the Sivers effect\cite{Sivers}, which is an initial 
state correlation between parton $k_T$ and the transverse spin of the nucleon. 
The second is transverse polarization of the quark, or transversity, and the
Collins effect\cite{Collins} in the fragmentation process.
The third involves a correlation between $k_T$ and quark spin within the 
unpolarized nucleon, which is believed to be very small\cite{Daniel}.  
Calculations of twist-3 contributions\cite{QS,koike} also have 3 terms 
similar to the models with transverse momentum.
While all these mechanisms can contribute at the same time, recent 
study\cite{nocollins} shows that the Collins effect is suppressed due 
to cancellations in quantum phases. 


\begin{figure}
  \centerline {\includegraphics[width=8.0cm,height=4.0cm,clip]{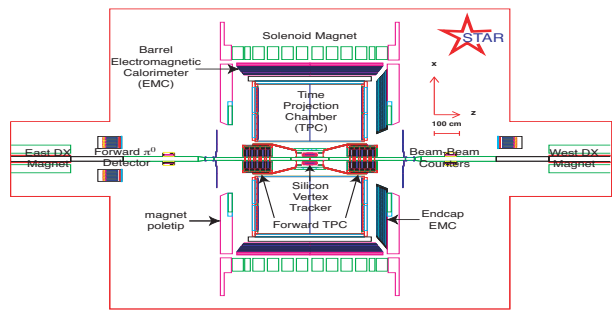}}
  \caption{\label{fig:star} Top view of the STAR detector. 
   }
\end{figure}

The Solenoid Tracker at RHIC (STAR) is a multipurpose detector at
Brookhaven National Laboratory.  One of its principle components is a
time projection chamber (TPC and Forward-TPC) embedded in a 0.5 T 
solenoidal magnetic field used for tracking charged particles 
at $|\eta|<1.2$ and $2.8<|\eta|<3.8$.
To extend the rapidity coverage, a forward $\pi^0$ detector (FPD) was
installed at STAR, as shown in Fig.\ref{fig:star}. The FPD is a 
7$\times$7 array of 3.8 cm$\times$3.8 cm$\times$45 cm Pb-glass cells
with individual photomultiplier tubes.
For the 2003 RHIC run, both beam left and right detectors on 
the east side were installed, and a beam left detector was installed
on the west side.  The FPD provides triggering and reconstruction of 
neutral pions produced with $3.3< \eta < 4.2$. 

The energy calibration of the FPD is done using reconstructed $\pi^0$ mesons.
The reconstruction method was extensively studied using GEANT 
Monte-Calro simulations\cite{geant}.
The absolute energy scale is known to better than $2\%$, and the reconstruction
efficiency is found to be mostly determined by geometry.
Photon conversion events primarily in the beam pipe producing hits
in both the Forward-TPC and the FPD were used to determine the position 
of the FPD.


\begin{figure}[ht]
\includegraphics[width=5.5cm,clip]{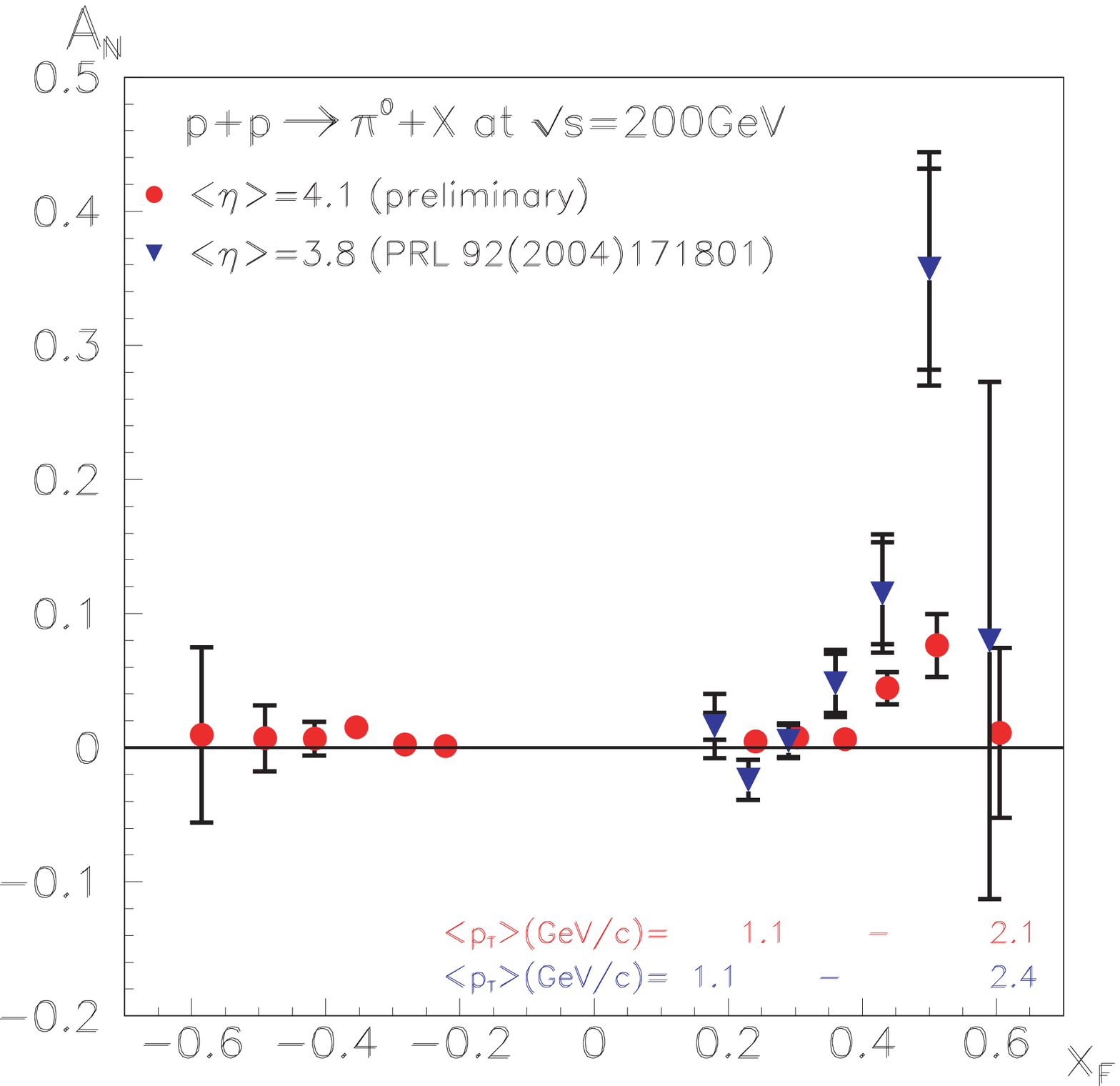}
\includegraphics[width=5.5cm,clip]{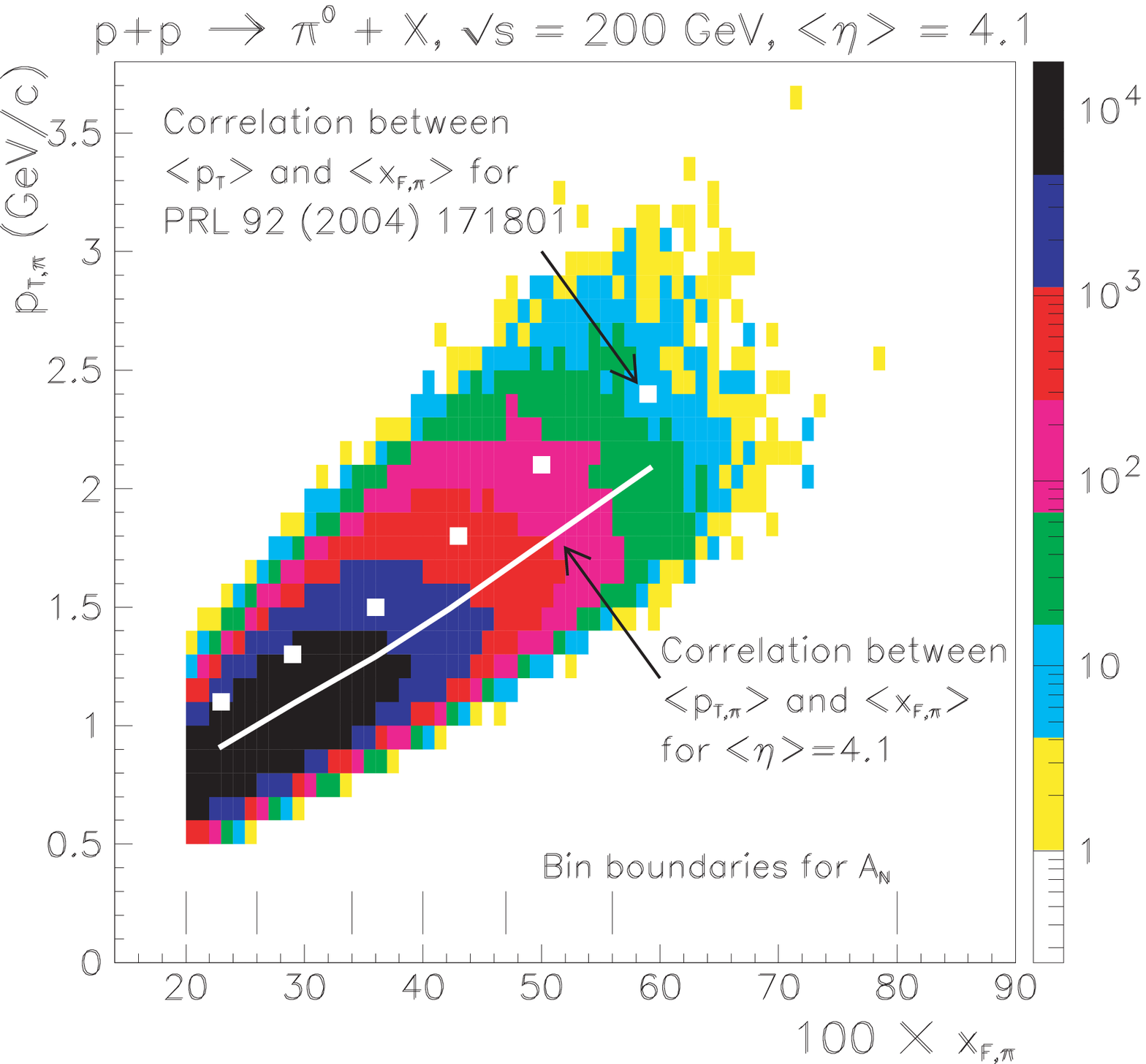}
\caption{ \label{fig:an}
Left: Preliminary results on analyzing power of $p+p \rightarrow \pi^0 + X$ 
as function of $x_F$ at $<\eta>=4.1$ (circle), compared to published 
results at $<\eta>=3.8$ (triangle). 
Right: $x_F$ and $p_T$ range of the data. The line shows the correlation
between $<x_F>$ and $<p_T>$ for $<\eta>=4.1$. The square dots shows the
$<x_F>$ and $<p_T>$ for $<\eta>=3.8$ 
.}
\end{figure}

In the 2003 RHIC run, p+p collisions were studied at $\sqrt{s}=200$ GeV
with average polarization $\sim25\%$ and integrated luminosity of $\sim 0.5/{\it pb}$. 
The polarization was measured by the pC CNI polarimeter\cite{osamu}. 
For the east side, the cross ratio method is used to obtain $A_N$. 
For the west side, the $A_N$ is obtained from spin dependent yields 
normalized by the Beam-Beam Counter\cite{joanna}. 
These 2 measurements are found to be consistent, and are combined. 
Positive (negative) $x_F$ is defined when the $\pi^0$ is observed with the same 
(opposite) longitudinal momentum as the polarized beam.
Positive $A_N$ is defined as more $\pi^0$ going left of the upward polarized 
beam for both positive and negative $x_F$. 

The preliminary results are shown in Fig.\ref{fig:an}.
The $A_N$ for positive $x_F$ at $<\eta>=4.1$ is found to be significantly 
non zero and positive, confirming earlier measurements.
The first measurement of $A_N$ at negative $x_F$ has been done,
and found to be consistent with zero. The negative $x_F$ results may 
give an upper limit on the gluon Sivers function\cite{nocollins}, 
which is currently unknown.

Future studies using the STAR FPD will include 
measurements of the $p_T$ dependence of $A_N$ at fixed $x_F$,
di-hadron correlations to distinguish the Collins effect from the Sivers effect 
and $A_{LL}$ which is sensitive to gluon polarization at low-$x$.

\end{document}